\documentclass{JHEP3}
\usepackage{latexsym}
\usepackage{mathrsfs}
\usepackage{amssymb}
\newcommand{\half}{\frac{\scriptstyle 1}{\scriptstyle 2}}
\newcommand{\C}{\mathbb{C}}
\newcommand{\CoM}{\mathbb{CM}}

\newcommand{\CP}{\mathbb{CP}}

\newcommand{\PT}{\mathbb{PT}}

\newcommand{\R}{\mathbb{R}}

\newcommand{\J}{\mathcal{J}}

\newcommand{\cPT}{\mathscr{PT}}
\newcommand{\M}{\mathbb{M}}
\newcommand{\CM}{\mathscr{M}}

\newcommand{\T}{\mathbb{T}}

\newcommand{\p}{\partial}
\newcommand{\dbar}{\bar\partial}
\renewcommand{\d}{\mathrm{d}}
\newcommand{\e}{\mathrm{e}}

\newcommand{\D}{\mathrm{D}}

\newcommand{\cA}{\mathcal{A}}
\renewcommand{\O}{\mathcal{O}}
\newcommand{\cT}{\mathscr{T}}

\newcommand{\End}{\mathrm{End}}
\newcommand{\SU}{\, \mathrm{SU}}
\newcommand{\GL}{\mathrm{GL}}
\newcommand{\Det}{{\mathrm{Det}}}
\newcommand{\Lie}{{\mathcal{L}}}

\newcommand{\tr}{\, \mathrm{tr}}
\newcommand{\TS}{\mathrm{TS}}

\newcommand{\hook}{{\setlength{\unitlength}{11pt}
    \begin{picture}(.833,.8)
      \put(.15,.08){\line(1,0){.35}}
      \put(.5,.08){\line(0,1){.5}}
      \end{picture}}}

\newcommand{\be}{\begin{equation}\label}
\newcommand{\ee}{\end{equation}}
\newcommand{\bea}{\begin{eqnarray}\label}
\newcommand{\eea}{\end{eqnarray}}

\newtheorem{propn}{Proposition}[section]

\newtheorem{lemma}{Lemma}[section]

\title{Twistor actions for non-self-dual fields; a new foundation for
  twistor-string theory}

\author{ L.J.Mason \\ The Mathematical Institute, 24-29 St Giles,
  Oxford OX1 3LB, England\\lmason@maths.ox.ac.uk} \abstract{ Twistor
  space constructions and actions are given for full Yang-Mills and
  conformal gravity using almost complex structures that are not, in
  general, integrable.  These are used as the basis of a derivation of
  the twistor-string generating functionals for tree level
  perturbative scattering amplitudes of Yang-Mills and conformal
  gravity.  The derivation follows by expanding and resumming the
  classical approximation to the path integral obtained from the
  twistor action.  It provides a basis for exploring whether the
  equivalence can be made to extend beyond tree level and allows one
  to disentangle conformal supergravity modes and super Yang-Mills
  modes from the standard Yang-Mills modes.  }

\preprint{July 27, 2005}

\begin{document}

\section{Introduction}
Twistor-string theory provides a dramatic reformulation of
perturbative $N=4$ super-Yang-Mills and conformal super-gravity
scattering amplitudes in terms of integrals over moduli spaces of
algebraic curves in super-twistor space (a supersymmetric version
$\CP^{3|4}$ of complex projective three space, $\CP^3$), Witten
(2004).  Whilst it is widely believed that the twistor-string
formulation is correct at tree level, no systematic proof is known.
The purpose of this article is to provide a derivation of these
formulae from first principles.  It starts with the space-time action,
and proceeds via a twistor space action associated to a corresponding
twistor construction for fields that are not necessarily self-dual.
The price we pay for this extra generality over and above the standard
twistor correspondences for self-dual fields is that the twistor
almost complex structures are no longer generally integrable.  This
limits the applicability of the constructions to problems in classical
geometry and reflects the lack of integrability of the classical
equations.  Nevertheless, the constructions are sufficent to provide a
derivation of twistor-string theory.  In particular we give a formal
proof that the twistor-string generating functionals for perturbative
scattering amplitudes are correct at the classical limit, i.e., for
tree diagrams; more work is required to extend the approach rigorously
to loop diagrams, although it provides a platform from which one can
investigate the problem. 
This approach also
disentangle Yang-Mills from conformal gravity and the supersymmetric
theories from their bosonic constituents, thus making the study of
loops much more straightforward; from the twistor-string point of
view, a higher genus contribution automatically includes conformal
supergravity modes and supersymmetric partners in the loops.
 The proof is formal to the extent that it
relies on an expansion of the classical limit of the path integral and
infrared divergences are not addressed.  

In twistor-string theory, scattering amplitudes for gluons in helicity
eigenstates are given by integrals over the moduli space of algebraic
curves in twistor space of degree $d$, where $d=q-1+l$, $l$ is the
number of loops and $q$ is the number of external gluons of helicity
$+1$.  The genus of the curves is also bounded by the number of loops.
Most of the investigations have been confined to tree diagrams and
hence are concerned with moduli spaces of rational curves (genus 0).
There have been, roughly speaking, two approaches to twistor-string
theory.  Cachazo, Svrcek and Witten (2004) consider integrals over the
moduli space of maximally disconnected curves, i.e., $d$ lines,
whereas Roiban, Spradlin and Volovich (2004) consider integrals over
moduli spaces of connected rational curves.  In the former approach,
the lines must be connected into a tree by holomorphic Chern-Simons
propagators although these are absent in the latter approach for tree
diagrams.  Gukov, Motl and Nietzke (2004) argue that the two
approaches are equivalent.

Perhaps the most elegant formula in the subject is that for the
on-shell generating functional for tree-level scattering amplitudes
$\cA[a,g]$ in the Roiban, Spradlin \& Volovich approach (here $(a,g)$
are the on-shell twistor fields, both being $(0,1)$-forms on a region
in $\CP^3$ with values in the endomorphisms of some given smooth
bundle $E\to\CP^{3}$, but with $g$ having homogeneity degree $-4 $ and
$a$ homogeneity degree 0).  In this case, the twistor on-shell fields
$(a,g)$ define a $\dbar$-operator $\dbar^s$ on a bundle $E$ over
$\CP^{3|4}$.  The generating functional for processes with $d+1$
external fields of helicity $+1$ is then: \be{deteq} \cA^d[a,g] =
\int_{\CM^d} \det (\dbar^s) \;\d \mu \ee where $\d \mu$ is a natural
measure on the moduli space $\CM^d$ of connected rational curves in
$\CP^{3|4}$ of degree $d$.\footnote{see the transparencies from
  Witten's lectures posted at www.maths.ox.ac.uk/$\sim$lmason/Tws.}
For the CSW version, extra terms associated with the holomorphic
Chern-Simons theory need to be incorporated also as described in
\S\ref{twistorstring} and this is the version that is proved here.  We
appeal to Gukov, Motl and Nietzke (2004) for the proof that this
implies the connected formulation given above.

Whilst the extensions to the $N=4$ supersymmetric versions of
Yang-Mills and conformal gravity are likely to be straightforward,
they are nevertheless complicated and ommitted here.  We work in
Euclidean signature throughout and ignore infrared divergences.

\medskip

\noindent
A summary of the rest of the article follows.

In \S\ref{action}, the Chalmer \& Siegel Lagrangian for the
anti-self-dual sector of Yang-Mills on space-time and its
generalisation to full Yang-Mills is set out.  Witten's twistor space
reformulation of the anti-self-dual sector as a holomorphic
Chern-Simons theory is then reviewed.  We then give a twistorial
formulation of the extra term, $I$, required in the action on twistor
space to generalise to full Yang-Mills.  This term is a two-point
integral on twistor space.  It is then shown that the full action
correctly reproduces full Yang-Mills theory on space-time by virtue of
a generalisation of the Ward construction for anti-self-dual gauge
fields to gauge fields that are not anti-self-dual.  The action is a
functional of a $\dbar$ operator on a bundle $E\to \PT$ where $\PT$ is
a region in $\CP^3$ and a homogeneous $(0,1)$-form $g$ with values in
$\End(E)$.  The construction relies on the fact that the restriction
of a $\dbar$-operator to a Riemann sphere, $\CP^1$, is automatically
integrable.  Not only does the twistor action reproduce the correct
equations of motion, but it also takes the same value as the
space-time action when evaluated on a solution to the field equations.
In \S\ref{construction} the twistor space action is expressed more
explicitly and the field equations derived and solved in terms
of an arbitrary solution to the Yang-Mills equations on space-time;
the general solution is gauge equivalent to such a solution.

\S\ref{twistorstring} contains the main derivation of the the
twistor-string on-shell generating functionals for tree-level
scattering amplitudes.  In \S\ref{susyaction} it is shown
that the the extra term $I$ has simple alternative expressions when
written in terms of integrals over super twistor space $\CP^{3|4}$.
Then in \S\ref{generatingfnls} the general definition of on-shell
generating functionals is reviewed and expressed as the classical
limit of a path integral.  In \S\ref{twistorstringgenfnls} the
twistor-string generating functionals are reviewed and that
appropriate to the Cachazo, Svrcek and Witten's approach is
presented.  This is then expanded and resummed to
show equivalence with the appropriate formulae from the twistor
Lagrangians derived in the previous section.  Since the classical
approximation uses only the value of the action and this takes
on the same value as the space-time action, this shows that the
twistor-string formulae provide the correct generating function for
Yang-Mills scattering theory at tree level.

In \S\ref{confgrav} the same process is worked through for conformal
gravity.  First we review the analogues of the Chalmer \& Siegel
Lagrangians appropriate to conformal gravity and their twistor space
reformulations as given by Berkovits and Witten (2004).  Then we move
on to finding the extra (non-local) term required in the twistor
action to extend to the full theory.  Finally in analogy with the
Yang-Mills case, we expand and resum the path integral to obtain the
relevant twistor-string formulae for the generating function for
perturbative scattering amplitudes.


\subsection*{Acknowledgements}
I would like to thank the Department of Mathematics at the University
of Edinburgh for hospitality while this work was completed.  I would
also like to thank Roger Penrose, Michael Singer and David Skinner for
a number of helpful remarks, and Philip Candelas, Xenia de la Ossa and
the speakers at the twistor-string workshop in Oxford, January
2005,\footnote{See www.maths.ox.ac.uk/$\sim$lmason/Tws for the
  collected slides from the talks.} for educating me about some of the
many facets of string theory, perturbative gauge theory and
twistor-string theory.

\section{The Yang-Mills actions}\label{action} 
For the purposes of this paper, we will Wick rotate to euclidean
signature and use the appropriate euclidean signature conventions.  It
is not clear that euclidean signature is essential for all of what
follows, but it helps avoid a number of technical difficulties.  Thus $\M$ will
denote $\R^4$ but with the standard flat euclidean metric $\eta$.  We will
take coordinates $x^a$, $a=0,\ldots 3$ on $\M$ and will use the metric
$\eta_{ab}$ to raise and lower indices as usual.  
We will denote self dual
spinors with a primed upper case roman index, e.g., $\pi_{A'}$,
$A'=0', 1'$.   Anti-self-dual spinors
will be denoted by $\omega^A$, $A=0,1$.  In Euclidean signature, the
reality structure is quaternionic
$$
\omega^A\rightarrow \hat\omega^A=(\bar \omega^1, -\bar\omega^0)
$$
so that $\hat{\hat\omega}^A=-\omega^A$.  We can
represent a vector
index as a pair of spinor indices, so the coordinates $x^a$ on $\M$
can be represented as $x^{AA'}$ and we define
$\p_{AA'}=\p/\p x^{AA'}$.

\subsection{The action on space-time}
The basic variable for Yang-Mills theory is a 1-form $A$ on Minkowski
space $\M$ with values in the Lie algebra of some gauge group.  
Let $F=\d A + [A,A]$ be the 
associated curvature; it is a Lie-algebra valued 2-form.  Then the
standard action for the Yang-Mills equations is
$$
S[A]=\int_\M \tr(F\wedge F^*)
$$ where $F^*$ is the Hodge dual of $F$ (in indices,
$F^*_{ab}=\half\varepsilon _{abcd}F^{cd}$) and $\tr$ is an ad-invariant
inner product on the Lie algebra.  Since $\int \tr (F^2)$ is a
topological invariant, for perturbative purposes, one can add any
multiple of this into the action without changing the perturbative
theory and this allows one to rewrite the action as
$$
S[A]=\int_\M\tr (F^+\wedge F^+)
$$ where $F^+=\half(F+F^*)$ is the self-dual part of $F$ satisfying
$F^{+*}=+F^+$.  (Here we have used the fact that $F=F^+ + F^-$ and
$F^+\wedge F^-=0$ automatically.)

The anti-self-dual sector of the theory is the case when $F^+=0$, but
in 
perturbing away from this to first order, one would want to introduce
$G$, a Lie algebra valued self-dual 2-form, so that $\epsilon G$
represents the infinitesimal value of $F^+$.
Chalmers \& Siegel have proposed the following action for the
anti-self-dual sector of the Yang-Mills equations on Minkowski space
$\M$:
$$
S_{\mathrm{asd}}[A,G]=\int_\M \tr(G\wedge F)
$$
The
Euler-Lagrange equations imply that $F$ is anti-self-dual, and $G$ is
covariantly closed.

To obtain the full Yang Mills equations we add the term
$-\frac \epsilon 2 I[G]$ where
$$
I[G]=\int_\M \tr (G\wedge G) \, .
$$
The action for full Yang-Mills is then
$$
S_{\mathrm{YM}}=S_{\mathrm{asd}}-\frac\epsilon 2I[G]\, .
$$
When $\epsilon\neq 0$, the Euler-Lagrange equations imply that
$F^+$, the self-dual part of $F$ is $\epsilon G$, which is in turn
covariantly closed, so that the full Yang-Mills equations are
satisfied. Forming a perturbation series in $\epsilon$ around
$\epsilon=0$ therefore gives a way of perturbing full Yang-Mills
theory around its anti-self-dual sector.

We also note that the value of this last action on a solution to the field
equations is, up to an overall multiplicative factor, the same as for
the standard Yang-Mills equations at least perturbatively so that the
topological term does not contribute.

\subsection{The actions on twistor space}
We first review twistor space geometry and notational conventions that
we will use.
Twistor space $\PT$ will be taken to be some neighbourhood of a line
in complex projective 3-space, $\CP^3$.  We will work in a space-time
of euclidean signature.  In this case,  we can choose our
neighbourhood so that $\PT$ fibres over an open set
$U\subset \M$,
$p:\PT \to U$, with fibre the Riemann sphere, $\CP^1$.  This fibre is best
thought of as the projectivisation of the space $\C^2$ of the 
self dual
spinors, $\pi_{A'}$, $A'=0', 1'$ at $x\in U$.  
Thus $(x^{AA'},\pi_{A'})$ are coordinates on non-projective twistor
space $\T$, and the projective space is obtained by modding out the
scale of $\pi_{A'}$.

Homogeneous coordinates on twistor
space are provided by $Z^\alpha=(\omega^A,\pi_{A'})$ where
$\omega^A=x^{AA'}\pi_{A'}$.  We note that the complex conjugation on
spinors induces a similar conjugation $Z^\alpha\to\hat Z^\alpha$ with
$\hat{\hat{Z}}{}^\alpha=-Z^\alpha$.  This conjugation restricts to give
the antipodal map on each $\CP^1$ fibre of $\PT\to\M$.
The coordinates $(\omega^A,\pi_{A'})$ are holomorphic coordinates for
the standard complex structure on
$\PT$.   In terms of these coordinates, the projection $p$ is given by
$$
p(\omega^A,\pi_{A'})=\{x^{AA'}=\frac1{\hat\pi^{C'}\pi_{C'}}\left( \omega^A
\hat\pi^{A'}-\hat\omega^A\pi^{A'}\right) \}\, .
$$
The complex structure can also be represented in terms of the
distribution of $(0,1)$ vectors $\D= \{
\p/\p\hat\pi_{A'},\pi^{A'}\p_{AA'}\}$, where $\p_{AA'}=\p/\p x^{AA'}$.
It can also be represented by 
the $\dbar$-operator (written here on the non-projective space)
$$
\dbar=\frac1{\pi^{A'}\hat\pi_{A'}}\d
x^{AA'}\hat\pi_{A'}\pi^{B'}\p_{AB'} +
\d\hat\pi_{A'}\frac{\p}{\p\hat\pi_{A'}} 
$$

The connection 1-form, $A$ is a connection on a bundle $E$ which, with
an abuse of notation, can be pulled back to give a smooth bundle
$E\to\PT$.  The connection then allows one to define a d-bar operator
$\dbar_a=\dbar +a$ on $E\to \PT$ where $a$ is a $(0,1)$-form with
values in $\End(E)$ and is the $(0,1)$-part of the pullback of $A$ to
$\PT$.  We will see that $G$ corresponds to a $(0,1)$-form $g$ with
values in $\End(E)\otimes\O(-4)$ where $\O(-1)$ is the tautological
bundle over $\CP^3$.  Witten (2004) shows that $S_{\mathrm{asd}}$ has
a direct analogue on twistor space in the form of the spin-1 part 
\be{twistasdaction}
S_{\mathrm{asd}} [a,g]=\int_\PT \tr(g\wedge f)\wedge \Omega
\ee
of a super Chern-Simons Lagrangian 
where $f:=\dbar a+a\wedge a=\dbar_a^2$ is the $(0,2)$-part of the
curvature of a connection with $(0,1)$-part $a$ and $\Omega =
\varepsilon_{\alpha\beta\gamma\delta} 
Z^\alpha\d Z^\beta \d Z^\gamma
\d Z^\delta\in \Gamma(\PT, \Omega^{(3,0)}(4))$ is the (weighted)
holomorphic volume form (here as usual $\varepsilon_{\alpha\beta\gamma\delta} =
\varepsilon_{[\alpha\beta\gamma\delta]}$, $\varepsilon_{0123}=1$).

The correspondence is precise for classical fields modulo their
appropriate gauge freedoms: the Euler-Lagrange equations from this
action imply that $f=0$ and $[\dbar+a,g]=0$.  The Lagrangian is
invariant under the usual group of gauge transformations
(automorphisms) of $E$ together with $g\rightarrow g+\dbar_a\chi$ for
smooth sections $\chi$ of $\End(E)(-4)$.  Thus, modulo gauge freedoms,
the first equation implies that $\dbar+a$ defines a holomorphic
structure on $E$ up to gauge transformations, and $g$ defines a
cohomology class in $H^1(U,\End(E)(-4))$.  Holomorphic vector bundles
$E$ correspond to anti-self-dual Yang-Mills gauge connections $A$ by
the Ward transform, and elements $g\in H^1(U,\End(E)(-4))$ corresponds
to covariantly closed self-dual 2-forms with values in the Lie algebra
of the gauge group by a standard generalisation of the Penrose
transform as follows.


In the abelian case, the Penrose tranform $g\to G$ is implemented by
\be{pentrg}G=p_* (g\wedge 
\Omega)\ee
that is, integrate
over the fibres of $p$ to obtain a 2-form on $\M$.
 This
necessarily provides a self-dual 2-form, since 
in
$(x^{AA'},\pi_{A'})$ coordinates we can write 
$$
\Omega= \D\pi\wedge \pi_{B'}\pi_{C'}\varepsilon_{BC}
\d
x^{BB'}\wedge\d x^{CC'}
\, , \qquad \mbox{ where }\quad
\D\pi=\pi^{A'} \d\pi_{A'} 
$$  
and so if we set
$$
G_{A'B'}(x)=\int_{L(x)}\pi_{A'}\pi_{B'}g\wedge\D\pi
$$
then the above formula gives
$$
G=G_{A'B'}\varepsilon_{AB}\d
x^{AA'}\wedge\d x^{BB'}\, ,
$$ which is necessarily self-dual.  It is easily seen that
$G$ must be closed since $g\wedge\Omega$ is.

In order to formulate full Yang-Mills on twistor space, we 
need to find the appropriate twistor version of the $I=\int_{\M} \tr
(G\wedge G)$ term.  
It follows from the above that in order to express $I[G]$ in terms of $g$
in the abelian case, we can consider the integral \be{Idef} I[g]=
\int_{\PT\times_\M\PT} \tr(g(Z_1)\wedge g(Z_2)) \wedge
\Omega(Z_1)\wedge\Omega(Z_2)\, , \ee where
$\PT\times_\M\PT=\{(Z_1,Z_2)\in\PT\times\PT | p(Z_1)=p(Z_2)\in\M\}$ is the
fibrewise product of $\PT$ with itself over $\M$ with fibre
$\CP^1\times\CP^1$.\footnote{This is the first point at which we needed to
have specified a real slice $\M$ of complex Minkowski space $\CoM$ (as
indeed one must for the ordinary action principle).  }


To make sense of (\ref{Idef}) in the nonabelian case, we must find
some way of comparing the fibre of $\End(E)$ at $Z_1$ with that at
$Z_2$.  In the integral we have already restricted to $\PT\times_\M\PT$
and so $Z_1$ and $Z_2$ both lie on the line $L(x)$ where
$x=p(Z_1)=p(Z_2)$.  Although the $\dbar_a$ operator is not a-priori
integrable, it is nevertheless necessarily integrable on restriction
to lines.  We make the assumption that $(E,\dbar_a)$ is
holomorphically trivial along such lines $L(x)$ for $x\in\M$; this
will be the case for small $a$ and hence perturbatively.  We
therefore define $\tr_a(g(Z_1)\wedge g(Z_2))$ to be the trace taken in
such a frame that is globally holomorphic along the line from $Z_1$ to
$Z_2$.  We now generalize equation (\ref{Idef}) to the non-abelian
case as
\be{Iadef} I[g,a]= \int_{\PT\times_\M\PT} \tr_a(g(Z_1)\wedge g(Z_2))
\wedge \Omega(Z_1)\wedge\Omega(Z_2)\, .  \ee This defines the
appropriate additional term, but is now a functional of $a$ also,
$I:=I[g,a]$.

This is not the most helpful form of $I[g,a]$ and we rewrite it as
follows.  First note that $\PT\times_\M\PT=\M\times\CP^1\times\CP^1$
and we coordinatize it by $(x,\pi_1,\pi_2)$ by setting
$$
(Z_1,Z_2)=\left( (x^{AA'}\pi_{1A'},\pi_{1A'}),
  (x^{BB'}\pi_{2B'},\pi_{2B'})\right)\, .$$
In these coordinates
\be{Omegaeqs}
\Omega(Z_1)\wedge \Omega(Z_2)= 
(\pi_{1}\cdot\pi_{2})^2 \D \pi_1\wedge \D\pi_2 \,
\d^4x 
\ee
where 
$$ \D\pi= \pi^{B'}\d\pi_{B'}\, , \quad \mbox{ and } \quad 
\pi_1\cdot\pi_2= \pi_1^{A'}\pi_{2A'}\, .
$$
Thus
$$
I[g,a]=\int_{\PT\times_\M \PT} \tr_a(g(Z_1)\wedge
g(Z_2))(\pi_1\cdot\pi_2)^2 \D\pi_1\wedge\D\pi_2\wedge \d^4x\, .
$$

In the non-abelian case, the integral formula for $G$ in terms of $g$
is: \be{Gintformula} G_{A'B'}(x)=\int_{\omega^A=ix^{AA'}\pi_{A'}}
\pi_{A'}\pi_{B'} g\wedge \pi_{A'}\d\pi^{A'} \ee as before, but the
integral must be performed in a holomorphic trivialisation of $E$ over
the Riemann sphere $p^{-1}(x)$.  The 2-form
$$
G=G_{A'B'}\varepsilon_{AB}\d x^{AA'}\wedge\d x^{BB'}=\int_{p^{-1}(x)}
g\wedge \Omega
$$
with the same proviso concerning the frame for $E$ and so we can
see that $I[g,a]=I[G]$.

We will therefore consider the twistor action
$$
S_T=S_{\mathrm{asd}}[a,g]-\frac\epsilon2 I[g,a].
$$
The gauge symmetry of this action is the group of gauge
transformations of the bundle $E\rightarrow \PT$ together with
$g\rightarrow g+\dbar_a\chi$.  It is easily seen that the action is
invariant under the group of gauge transformations of $E$.  To see
invariance under $g\rightarrow g+\dbar_a\chi$ for $I[g,a]$, note that
in the frame that is holomorphic up the fibres of $p$ in which the
trace is taken, $\dbar_a=\dbar$ on restriction to the fibres of $p$
and so the integral over one of the $\CP^1$ factors of a fibre will
give zero if we replace the corresponding $g$ by $\dbar\chi$.  The
invariance of $S_{\mathrm{asd}}$ is elementary.

\begin{propn}
The action $S_T=S_{\mathrm{asd}}[a,g]- \frac \epsilon 2 I[a,g]$ is
equivalent at the classical level to $S_{\mathrm{YM}}$.  This is true
both in the sense that gauge equivalence classes of solutions to
the Euler Lagrange equations on twistor space are in $1:1$
correspondence with gauge equivalence classes of solutions to the
Yang-Mills equations on space-time, and in the sense that the
twistor action takes the same values on $(a,g)$ as the space-time
action does on the corresponding $(A,G)$.
\end{propn}

\noindent
{\bf Proof:} In this subsection we give a quick but inexplicit proof
of this theorem.  In the next we will develop more notation so as to
be more explicit.

Given $(E\to\PT, a,g)$ satisfying the variational equations of the
action $S_T$, we wish to construct $(E\to\M, A,G) $ satisfying the
Yang-Mills equations.  We first define the bundle $E\rightarrow\M$ to
be the bundle whose fibre at $x$ is the space of global
$\dbar_a$-holomorphic sections of $E\to p^{-1}(x)\subset\PT$
(recalling that the bundle is assumed to be trivial on such lines).

We then note that we can define $G\in \Omega^{2+}\otimes \End (E)$ at
each $x$ to be the two form obtained by pushing down $g\wedge \Omega$
to $\M$ in the associated global holomorphic frame of $\End(E)$ over
$p^{-1}(x)$.  This necessarily provides a self-dual 2-form as before
and we see that
$$
I[g,a]=I[G]\, .
$$

The classical equations of motion obtained by varying $g$ are
\be{fieldeq} \dbar a + a\wedge a= \epsilon \int_{Z'\in p^{-1}(p(Z)) }
g(Z')\Omega(Z')\, .  \ee where the left hand side is evaluated at
$Z$ and as usual the integration is in a global
holomorphic trivialisation of $E$ over $p^{-1}(p(Z))$.  The integral
therefore yields the projection $G_{(0,2)}$ of $p^*G(p(Z))$ onto the
$(0,2)$-forms at $Z$. Thus 
$$\dbar_a^2=G_{(0,2)}\, .
$$

It follows that $f=\dbar_a^2$ has no component up the fibres of $p$.
This allows us to define a connection $A$ on $E\to\M$ as follows.
Pull back a section $s$ of $E\to \M$ to $E\to \PT$, then $\dbar_a s$
is holomorphic up the fibre of $p:\PT\to\M$.  More concretely on
$\PT$, $\pi^{A'}\p_{AA'}\hook \dbar_a s$ is therefore holomorphic
in $\pi_{A'}$. It is also global with homogeneity degree 1 over the
Riemann sphere with homogeneous coordinates $\pi_{A'}$, and depends
linearly on $s$.  We can deduce from a generalization of Liouville's
theorem that $\pi^{A'}\p_{AA'}\hook\dbar_a s=\pi^{A'}(\p_{AA'}
+A_{AA'})s$ for some connection 1-form $A=A_{AA'}\d x^{AA'}$ on $\M$.
This is in effect the standard Ward argument for constructing a
connection $A$ on $E\to\M$ from $a$.  The $\dbar_a$ operator can
therefore be represented in a frame pulled back from $E\to\M$ as
$$
\dbar_a=\frac{1}{\pi^{C'}\hat\pi_{C'}}
\d x^{AB'}\hat\pi_{B'}\pi^{A'}(\p_{AA'}+A_{AA'})
+\d\hat\pi_{C'}\frac{\p}{\p\hat\pi_{C'}}\, .
$$

Now we claim that
$S_{\mathrm{asd}}[a,g]=S_{\mathrm{asd}}[A,G]$.  This follows by using
the gauge invariance on twistor space to use a gauge pulled back from
$E\to\M$.  In this gauge, $a$ is the projection onto $(0,1)$-forms of
$A_{AA'}\d x^{AA'}$ and $f$ is the projection onto $(0,2)$-forms of
$F^+$.  Thus $S_{\mathrm{asd}}[a,g]=\int \tr (F^+\wedge
g)\wedge\Omega$.  Integrating over the fibres of $p$ then gives
directly that $S_{\mathrm{asd}}[a,g]=S_{\mathrm{asd}}[A,G]$.

We have now reduced the desciption to that of the Chalmers \& Siegel
Lagrangian and so we have obtained the appropriate field equations as
claimed.    $\Box$

\medskip

We have not in fact proved everything here: we have only provided a
map from solutions to the field equations associated to $S_T$ to those
of $S_{\mathrm{YM}}$.  To see that it is $1:1$ and onto gauge
equivalence classes we need to work more explicitly which we do in
the next subsection.

\subsection{The twistor action, field equations and solutions}
\label{construction} 

We can therefore take the full twistor-space Lagrangian to be
$$
S_T[a,g]=S_{\mathrm{asd}}
[a,g]-\frac\epsilon 2 I[g,a]\, .
$$
The equation of motion obtained by varying $g$ is given in equation
(\ref{fieldeq}) but that obtained by varying $a$ is more complicated,
and we now make the $I[g,a]$ term more explicit in order to calculate
the equations of motion.  We also show in this subsection how every
solution to the Yang-Mills equations on space-time gives rise to a
solution to the Euler-Lagrange equations of the twistor action and
that every solution to the Euler-Lagrange equations of the twistor
action is gauge equivalent to a space-time solution arising in this
way.

In the following our expressions will be functions of two or more
twistors, $Z_1$, $Z_2$, $\ldots$ or $\pi$ spinors, $\pi_1$, $\pi_2$,
\ldots.  We will adopt the convention that $g_1$ will denote a
function of $Z_1$ and so on.

We first introduce global holomorphic frames $F(x,\pi)$ over the line
$L_x$ corresponding to $x\in\M$ by $F(x,\pi):E_Z\rightarrow \C^r$,
($r$ is the rank of the bundle $E$) where $\dbar_a F|_{L_x}=0$ and
$F(x,\pi)$ is unique up to $F(x,\pi)\rightarrow F(x,\pi)\gamma(x)$
where $\gamma$ is a function on $\M$ with values in the gauge group.
We can then write $I[g,a]$ as
$$
I[g,a]=\int_{\M\times\CP^1\times\CP^1} \tr(
F_1^{-1}g_1F_1\wedge F_2^{-1} g_2F_2)\wedge \Omega_1\wedge\Omega_2\, ,
$$
where $F_1$ and $g_1$ are evaluated at $\pi_1$ and $Z_1$ with
$Z_1=(x^{AA'}\pi_{1A'},\pi_{1A'})$, etc..  

To reformulate this further, we note that the greens function
$K_{12}:=K(x,\pi_1,\pi_2)$ for the d-bar operator $\dbar_a|_{L_x}$ on
sections of $E\otimes\O(-1)|_{L_x}$ is, for $Z_1, Z_2\in L_x $,
$$
K_{12}=\frac 1{2\pi i}\frac{F_1F_2^{-1}}{\pi_1\cdot\pi_2}
$$
thus using equation (\ref{Omegaeqs}) and ignoring certain multiples
of $2\pi i$ (which can be absorbed into the definition of $\epsilon$
we can put
\be{newI} 
I[g,a]=\int_{\M\times \CP^1\times\CP^1} \tr(K_{21} g_1 K_{12}g_2)
(\pi_1\cdot\pi_2)^4\D\pi_1\D\pi_2\;\d^4 x
\ee where $D\pi=\pi^{A'}\d\pi_{A'}$ and we use the
fact that $\Omega_1\wedge\Omega_2=(\pi_1\cdot\pi_2)^2\d^4x\wedge
D\pi_1\wedge\D\pi_2$ as above.

The variation of $K$ with respect to $a$ is given by
$$
\delta K_{12}=\int
K_{13}\delta a_3 K_{32}\; \D\pi_3  \, .
$$
We can use this to calculate the variation of $I[g,a]$ with respect to
$a$  and hence the Euler-Lagrange equation obtained by varying $a$ in
the action.  This yields, after some manipulation,
\be{eqmotion2}
\dbar_{a_3} g_3=\epsilon 
\int_{\CP_1\times\CP_1} [K_{31} g_1 K_{13},
K_{32} g_2 K_{23}](\pi_1\cdot\pi_2)^3\pi_{1(A'}\pi_{2B')}\;
\D\pi_1\D\pi_2 \; \d^2x^{A'B'}_{(0,2)}\, .
\ee
In this notation, the equation of motion from varying
$g$ (\ref{fieldeq}) is
\be{eqmotion1}
\dbar a_1+a_1\wedge a_1=\epsilon \int_{\CP_1} K_{12}g_2 K_{21}
(\pi_1\cdot\pi_2)^2\pi_{2A'}\pi_{2B'}\D\pi_2\; \d^2x^{A'B'}_{(0,2)}\, .
\ee
where in both the above two equations $\d^2x^{A'B'}_{(0,2)}$ denotes
the $(0,2)$-part of $\d^2x^{A'B'}:=\varepsilon_{AB}\d x^{AA'}\wedge\d
x^{BB'}$ which is $\d^2x^{A'B'}_{(0,2)} = \d^2x^{C'D'}
\hat\pi_{C'}\hat\pi_{D'}\pi^{A'}\pi^{B'}/(\pi\cdot\hat\pi)^2$.   

As a check on these equations, it is helpful to see how they can be
solved in terms of the standard space-time data for a solution to the
full Yang-Mills equations.  Thus, let $A$ be a connection 1-form on
$\M$ for a solution to the full Yang-Mills equations and let
$G_{A'B'}$ be the self-dual part of its curvature.  The 
space-time field
equations are 
$$
\p^A_{(A'}A_{B')A}+ A_{(A'}^AA_{B')A}=\epsilon G_{A'B'}\, , \qquad
\nabla^{A'}_AG_{A'B'} =0\, ,
$$
where $\nabla_{AA'}=\p_{AA'}+A_{AA'}$ is the gauge covariant
derivative and is understood to act in the standard way on the adjoint
representation. 
Using the standard Euclidean fibration $p: \PT\rightarrow\M$, we
define $a$ to be the $(0,1)$ part of the pull-back of $A$ to
$\PT$ and using Woodhouse (1985), we define
\begin{eqnarray}\label{harmgauge}
a&=&\frac1{\pi\cdot\hat\pi}A_{AA'}\pi^{A'}\hat\pi_{B'}\d x^{AB'}\, ,
\nonumber \\ g&=&\frac1{(\pi\cdot\hat\pi)^4} \left( 3G_{A'B'}\hat\pi^{A'}
  \hat\pi^{B'} \D\hat\pi + \nabla_{AA'}G_{B'C'}\hat\pi^{A'}
  \hat\pi^{B'} 
  \hat\pi^{C'}\hat\pi_{D'} \d x^{AD'}\right) \, .
\end{eqnarray}
It can now be checked that if the Yang-Mills equations hold, we have,
in this gauge,
\begin{eqnarray*}
\dbar a+ a\wedge a&=& \epsilon G_{A'B'}\pi^{A'}\pi^{B'}\d^2x_{(0,2)}
\\ \dbar_a g&=&\frac\epsilon{(\pi\cdot\hat\pi)^2} \left( 
  \hat\pi^{A'}\hat\pi^{B'}[ G_{A'}^{E'},G_{B'E'}]
  \d^2 x_{(0,2)}\right) \, ,
\end{eqnarray*}
where $\d^2x_{(0,2)}$ is the $(0,2)$-form with values in $\O(-2)$
$$
\d^2x_{(0,2)}:=\frac1{(\pi\cdot\hat\pi)^2}\varepsilon_{AB}
\hat\pi_{A'}\hat\pi_{B'}\d  
x^{AA'}\wedge\d x^{BB'}
$$
In this gauge, the matrix $F$ can be taken to be the identity,  and
$K_{12}=1/2\pi i (\pi_1\cdot\pi_2)$ and equations (\ref{eqmotion1})
and (\ref{eqmotion2}) can be verified using equation
(\ref{Gintformula}). 

We finally note that since both the action
and field equations are invariant under the full group of gauge
transformations on twistor space 
$$
(\dbar_a,g)\rightarrow (\dbar_{a'},g')=(H^{-1}\dbar_a H,
H^{-1}(g+\dbar_a \chi) H)$$
where $H$ is an arbitrary smooth complex gauge transformation of the
bundle $E \rightarrow\PT$ and $\chi$ an arbitrary smooth section of
$E\otimes\O(-4)$ over $\PT$, given an arbitrary solution to equations
(\ref{eqmotion1},\ref{eqmotion2}), we can find a gauge transformation
to a frame that is holomorphic on the fibres of $p:\PT\rightarrow \M$
(so that $a$ vanishes on restriction to the fibres of $p$) and so that
$g$ is a harmonic representative on each of the fibres of $p$, see
Woodhouse (1985).  If,
furthermore, $(a,g)$ are solutions to
(\ref{eqmotion1},\ref{eqmotion2}) then 
we know from the previous subsection that they correspond to a
solution of the Yang-Mills equations and that $a$ and the vertical part of
$g$ have the form given above.  It is then straightforward to see that
the solution is precisely as given above up to a space-time gauge
transformation. 

We therefore see that the solutions to the Euler-Lagrange equations of
the twistor action are in $1:1$ correspondence with gauge equivalence
classes of solutions to the space-time Yang-Mills equations.

\section{Twistor-string Yang-Mills generating
  functionals from the twistor action}\label{twistorstring} 

The twistor-string formulae refer to a holomorphic Chern-Simons theory
on super twistor space $\PT_s$ which is an appropriate subset of
$\CP^{3|4}$.  This space is obtained by introducing odd homogeneous
twistor coordinates $\psi_i$, $i=1,\ldots,4$ in addition to the
standard bosonic homogeneous coordinates $Z^\alpha$ so that
$\CP^{3|4}$ is the space of non-zero $(Z^\alpha,\psi_i)\in\C^{4|4}$
modulo the equivalence relation $(Z^\alpha,\psi_i)\sim (\lambda
Z^\alpha,\lambda\psi_i), \lambda\in\C^*$.


Chiral Super-Minkowski space $\M_s$ is then $\R^{4|8}$ with
coordinates $(x^{AA'},\theta^{iA'})$ and there is the incidence
relation with supertwistor space given by 
\be{superincidence}
(\omega^A,\pi_{A'},
\psi_i)=(x^{AA'}\pi_{A'}, \pi_{A'}, \theta^{A'}_i\pi_{A'})\, .
\ee
Since we have stripped out all the superpartners except those of
helicity $\pm 1$, we set $\Psi=\psi_1\psi_2\psi_3\psi_4$ and this will
be the supersymmetric quantity that we will use most. 

In the following our expressions will be functions of two or more
twistors, $Z_1$, $Z_2$, $\ldots$ or $\pi$ spinors, $\pi_1$, $\pi_2$,
\ldots.  We will again adopt the convention that $g_1$ will denote a
function of $Z_1$ and so on.

\subsection{Supersymmetric D-instanton reformulation of twistor
  action}\label{susyaction} To make closer contact with twistor string
formulae, we can consider the $(\pi_1\cdot\pi_2)^4$ term in equations
(\ref{newI})to arise from a superspace integral using the identity
$$
\int \Psi_1\Psi_2\; \d^8\theta=(\pi_1\cdot\pi_2)^4\, ,
$$ where $\Psi_1=\Pi_{i=1}^4 \theta_i^{A'}\pi_{1A'}$ and similarly for
$\Psi_2$.  Therefore we can write:
$$
I[g,a]=
\int_{\M_s}\int_{L_{(x,\theta)}
\times L_{(x,\theta)}} \tr(K_{21} \Psi_1g_1 K_{12}\Psi_2g_2)
\; \D\pi_1\D\pi_2\, \d^4 x\, \d^8\theta
$$

We now wish to reformulate the Green's functions $K_{12}$ in terms of
vacuum expectation values of fermion currents.  We use a device
introduced in Mason, Singer \& Woodhouse (2002) in the context of the
Ward construction for integrable systems. Introduce fermion spinor
fields $\alpha$ and $\beta$ (i.e., fields of homogeneity $-1$) on each
$L_x$ taking values in $E$ and $E^*$ respectively with action
$$S[\alpha,\beta]=\int_{L_{(x,\theta)}}\beta\dbar_a\alpha\wedge\D\pi
$$  where $L(x,\theta)$ is the line in super twistor space given by
holding $(x,\theta)$ fixed in equation (\ref{superincidence}).
These will be the D-instantons of twistor-string theory.
Then
$$
K_{12}=\langle\alpha_1\beta_2\rangle 
$$
where $\langle \O\rangle$ denotes the vacuum expectation of the
operator $\O$ associated to the quantum field theory of the fermions
$\alpha $ and $\beta$ on $\CP^1$.  In the above we are taking
$\alpha_1$ and $\beta_2$ to be associated to the same line,
$L(x,\theta)$; if they are taken to be associated to $L(x,\theta)$ and
$L(x',\theta')$ respectively, we would have instead $K_{12}$ if
$(x,\theta)=(x',\theta')$
or zero otherwise---there are no singular terms in $x$.  With this,
we can express $I[g,a]$ as follows:
\begin{eqnarray}
I[g,a]&=&\int_{\M_s} 
\left\langle\int_{L_{(x,\theta)}\times L_{(x,\theta)}}
\tr(J_{a1} \Psi_1
g_1)\tr(J_{a2} \Psi_2 g_2)
\right\rangle \d^4x\d^8\theta \nonumber \\
&=&\int_{\M_s} \d^4x\d^8\theta \left\langle \left(\int_{L_{(x,\theta)}}
\tr(J_{a} \Psi
g)\right)^2
\right\rangle \label{YMfin}
\end{eqnarray}
where $J_a=:\alpha\beta:\D\pi$ is the current associated to $\alpha$
and  
$\beta$ (the :: denoting wick-ordering and the subscript $a$ denotes
dependence on $a$) and the subscript 1 or 2
denotes evaluation at a point of the first or second factor of
$L(x,\theta)\times L(x,\theta)$.

\subsection{A digression concerning generating functions for scattering
amplitudes}\label{generatingfnls}

We now wish to use this twistor form of the Yang-Mills action to show
that the conjectured formulae for the generating functional
$\cA_{\mathrm{TS}}[a,g]$ for tree level QCD amplitudes from twistor
string theory is the same as the generating functional
$\cA_{\mathrm{YM}}$ obtained from the standard Yang-Mills action.
We first need to establish some basic facts about 
these generating 
functions.

We are concerned here with on-shell generating functionals rather than
the more common off-shell generating functional, usually denoted
$Z[J]$, which is a functional of a source usually denoted $J$ which is an
arbitrary function on space-time.  Instead, both the twistor-string
and and the standard Yang-Mills generating functionals $\cA[a,g] $ are
functionals of linearised free gluon fields, which we will represent
by their on-shell twistor data $(a,g)$ where on shell means that they
satisfy the linearised equations at $\epsilon=0$: $a$ and $g$ are
therefore both linear dolbeault cohomology classes of homogeneity $0$
and $-4$ respectively.

The $\cA[a,g]$ are generating functionals in the sense
that n-point scattering amplitudes are obtained as
functionals of positive frequency linear fields by taking the n'th
functional derivative of $\cA_{\mathrm{TS}}$ with respect to $(a,g)$
in the directions of the given positive frequency linear fields and
evaluating at $(a,g)=(0,0)$.  The fact that we are considering a
functional only of positive frequency fields means that the generating
functional generates diagrams with no incoming fields, just outgoing
fields.  This is sufficient as crossing symmetry then allows one to
construct all other processes.

For a generic quantum field theory of a field theory for a field
$\phi$ with action, say $S[\phi]=\half(\p_a\phi\p^a\phi -
m^2\phi^2)-\lambda V(\phi)$, such a generating functional would have
the path-integral expression
$$
\cA[\phi]=\int \D\tilde \phi \;\exp \frac i\hbar S[\tilde \phi]
$$
where the functional integration is understood to be over fields
$\tilde \phi$ such that, as $t\rightarrow +\infty$, the negative
frequency part of $\tilde\phi-\phi$ tends to zero, and as
$t\rightarrow -\infty$, the positive frequency part of
$\tilde\phi-\phi$ tends to zero, Faddeev \& Slavnov (1991).
Thus $\cA[\phi]$ is the `wave functional' of the theory.

Given the complications associated with defining even the perturbation
series for a functional integral, it is difficult to deduce
rigourously that the quantum theory will be correctly reproduced after
a manipulation of the exponential of the action in the path integral.
However, more can be said about the classical limit.

The classical limit $\cA^{\mathrm{cl}}[\phi]$ generates all the tree
diagrams.  It can be obtained by first constructing the classical
solution $\tilde \phi$ that is appropriately asymptotic to $\phi$ as
above by iterating the integral form of the field equations to produce
a sequence of fields $\phi_n$ such that 
$$
\phi_{n+1}(x)= \phi(x)+ \lambda \int
\Delta_F(x,x')V(\phi_n(x'))\d^4x' 
$$
where $\Delta_F$ is the Feynman propagator that inverts $\p_a\p^a
+m^2$ and $\phi_0=\phi$.  We can then define
$\tilde\phi=\lim_{n\rightarrow\infty}\phi_n$ as a power series in
$\lambda$.  Then we have
$$\cA^{\mathrm{cl}}=\exp \frac i\hbar S[\tilde\phi].$$ 

In such perturbative studies, the free field $\phi$ is taken to be a
plane wave.  Such fields are entire on complex Minkowski space (but
are singular at infinity in the conformal compactification).  It
follows from the analyticity properties of the Feynman propagator that
the corresponding solution $\tilde\phi$ can be analytically continued
to the Euclidean section from the Minkowski signature section.  We can
therefore assume that all integrals are over the Euclidean section
$\M$.

Following the above, the generating functional for tree level
Yang-Mills amplitudes $\cA_{\mathrm{YM}}[a,g]$ is given by 
\be{YMgen1}
\cA_{\mathrm{YM}}[a,g]=\exp \frac i\hbar \left(S_{\mathrm{asd}}[\tilde
  a,\tilde g]+ \frac{\epsilon^2\hbar}{2i}I[\tilde g,\tilde a]\right)\,
, \ee 
where it is undertood that $(\tilde a,\tilde
g)=\lim_{n\rightarrow\infty}(a_n,g_n)$ where 
$$
(a_{n+1},g_{n+1})=(a,g) + \dbar^{-1}\left( -a_n\wedge a_n + \epsilon
  \ldots, 
-a_n\wedge g_n + \epsilon\ldots\right)
$$
where the $\ldots$ denotes the terms on the right hand sides of
equations (\ref{eqmotion1}) and (\ref{eqmotion2}).  

We deduce from this that any formal manipulation of the exponential of
the action that preserves the value it takes on solutions to the
equations will give rise to the correct tree diagrams in perturbation
theory.  This will nevertheless be somewhat formal as, even at tree
level, these series exhibit infrared divergences.  Infrared
divergences are a standard problem in quantum field theory with a
number of standard resolutions and we will not consider them further.

\subsection{The twistor string generating
  functionals}\label{twistorstringgenfnls} 
The conjectures for the twistor-string form of the generating
functional are only confident in the classical approximation
corresponding to tree diagrams in perturbation theory and rational
(genus 0) curves in twistor space; we restrict attention to this
classical limit here and omit the `cl' superscript on $\cA$ in
the following.  There is some evidence that the conjecture might be
valid for the full quantum field theory (which would correspond to
loops in perturbation theory and curves of higher genus in twistor
space in twistor-string theory) but we will not be able to say
anything definitive here.

In its simplest form, the generating functional is given by:
\be{tsgenerate} \cA_{\TS}[a,g] = \sum_d \int_{\CM^d_s}
\Det(\dbar_{a+\epsilon\Psi g}|_C)\;\d\mu \ee Here $C\in \CM^d_s$ where
$\CM^d_s$ is a totally real submanifold (or contour) in the space of
connected degree-$d$ rational curves in super-twistor space,
$\epsilon$ is a small parameter used to expand about the self-dual
sector of the theory and $\d\mu$ is a naturally defined measure on
$\CM^d_s$.  This determinant has a standard functional integral
representation
$$
\cA_\TS [a,g]=\sum_d \int_{\CM^d_s} \int\D\alpha\D\beta \exp
\left(\int_C 
\beta\dbar_{a+\epsilon\Psi g}\alpha \right)\, ,
$$
where the functional integral is over the space of $\alpha$s and
$\beta$s which are fermionic spinors on each $C$ with values in $E$
and $E^*$ respectively.  These conjectured forms are only confidently
expected to be valid for tree diagrams when we consider connected
rational curves and a special normal form of the Dolbeault
representatives for $(a,g)$.

Here we will only make contact with the MHV diagram formulation of
twistor-string theory, Cachazo, Svrcek and Witten (2004), in which
instead of connected rational curves, we consider maximally
disconnected rational curves so that each $C$ is the union of $d$
lines (degree 1 curves) in super-twistor space.  In this approach, the
$d$ lines need to be connected into a tree by Chern-Simons
propagators.  It has been argued that this disconnected formulation is
equivalent to the connected formulation by Gukov, Motl and Nietzke
(2004).

To obtain a generating functional in the MHV diagram formulation, we
write $C=\cup_{r=1}^d L(x_r,\theta_r)$ where $(x_r,\theta_r)$ are $d$
points in super Minkowski space $\M_s$.  The moduli space of such
disconnected curves is therefore the $d$-fold product of
super-Minkowski space $\M^d_s$.  The generating function for this
version of the theory needs to include the Chern-Simons action to give
$$ 
\cA_\TS[a,g]=
\e^{\frac i\hbar S_{\mathrm{asd}}[\tilde
a,\tilde g]}\sum_d \int_{\M^d_s}\d\mu_d\;
\int\D\alpha\D\beta\; \exp
\left(\sum_{r=1}^d\int_{L(x_r,\theta_r)} \beta\dbar_{\tilde
  a+\epsilon\Psi \tilde g}\alpha 
\right)\, ,
$$
where $\d\mu_d= \Pi_{r=1}^d\d^4 x_r\d^8\theta_r$ and here, as in
the previous subsection, $(\tilde a, \tilde g)$ are understood to be
the solutions to the classical field equations obtained by iterating
the appropriate integral versions of the field equations with
inhomogeneous terms given by $(a,g)$.\footnote{In the main
  applications of twistor string theory $(a,g)$ are taken to be the
  Penrose transform of plane waves which are entire on complex
  Minkowski space, but singular at infinity.  Hence, $(a,g)$ can be
  defined smoothly over the 
  complement of the line in $\PT$ corresponding to the point at
  infinity in space-time.  Wick rotation, using the
  analycity properties of the Feynman propagator and its counterpart
  on twistor space, can then be invoked
  to analytically continue the integrals over the Euclidean section.
  We will therefore assume that all integrals are over the Euclidean
  section $\M$.
}

Expanding this in $\epsilon$, 
it is straightforward to see that the supersymmetric integrals over
$\theta_r$ only give nontrivial
contributions from terms in the expansion in which
there are precisely two $\Psi$s integrated over each set of
$\theta_r$.  Thus $\cA_\TS=\sum_d \epsilon^{2d}\cA_\TS^d$ where
\begin{eqnarray}
\cA_\TS^d&=&
\e^{\frac i\hbar S_{\mathrm{asd}}[\tilde
a,\tilde g]} 
\int_{\M^d_s} \d\mu_d 
\int\D\alpha\D\beta \; \e^{
\left(\sum_r\int_{L_{(x_r,\theta_r)}} \beta\dbar_{\tilde
    a}\alpha\right)} 
\frac{(2d)!}{2^d d!}\frac{\Pi_s\left(\int_{L_{(x_s,\theta_s)}}\beta
    \Psi \tilde g\alpha 
\right)^2}{(2d)!} \nonumber\\
&=& \e^{\frac i\hbar S_{\mathrm{asd}}[\tilde
a,\tilde g]} 
\int_{\M^d_s} \Pi_{r=1}^d\d^4 x_r\d^8\theta_r
 \frac{\left\langle\left(\int_{L_{(x_r,\theta_r)}}\beta \Psi \tilde
       g\alpha 
\right)^2\right\rangle}{2^dd!} \nonumber\\
&=&\e^{\frac i\hbar S_{\mathrm{asd}}[\tilde
a,\tilde g]} \frac{I[\tilde g,\tilde a]^d}{2^d d!}
\, , \label{maincalc}
\end{eqnarray}
where the combinatorial factor in the first line comes from the number
of choices of pairs of integrals over the $i$th copy of $\M_s$ over
the $2d$ factors in the $2d$th term of the expansion of the
exponential and in the second we have used the formula for $I[g,a]$ in
equation (\ref{YMfin}).  We can now resum over $d$ to obtain
\be{YMgen} 
\cA_\TS=\exp \frac i\hbar \left(S_{\mathrm{asd}}[\tilde
    a,\tilde g]+ \frac{\epsilon^2\hbar}{2i}I[\tilde g,\tilde
    a]\right)\, .  
\ee 
As can be seen, up to a
redefinition of the expansion parameter, this gives rise to the
classical Yang-Mills action generating functional as desired.

\subsection{Extension to the full quantum field theory}
The full quantum field theoretic generating functionals for Yang-Mills
are expressed formally in terms of the path integral
$$
\cA[a,g]=\int \D \tilde a\D\tilde g \exp\frac i\hbar \left(
  S_{\mathrm{asd}}[\tilde a,\tilde g] + \epsilon I[\tilde g,\tilde
  a]\right) 
$$
It is clear that formally the expansion and resummation in
(\ref{maincalc})and (\ref{YMgen}) will be possible in the full
quantum field theoretic path integral as in the generating functionals
for tree diagrams discussed above to yield
$$
\cA[a,g]=\int \D \tilde a\D\tilde g \D\alpha\D\beta\;\e^{\frac i\hbar 
  S_{\mathrm{asd}}[\tilde a,\tilde g]} \sum_d\int_{\M_s^d}\d\mu_d\;
\e^{\sum_{r=1}^d\int_{L(x_r,\theta_r)}\beta\dbar_{\tilde a+ \epsilon
    \Psi\tilde g}\alpha}
$$
However, in the full quantum field theory we would need to consider
the gauge fixing.  There is a useful Poincar\'e invariant twistor
space gauge in which the $(0,1)$-forms $(a,g)$ on twistor space are
orthogonal to the fibres of the projection $\PT\to \CP^1$.  This is
particularly useful because it linearizes the Chern-Simons theory as
$a\wedge a$ and $a\wedge g$ vanish identically.  This is the gauge in
which most of the twistor-string calculations have so far taken place.
There is also a gauge that is adapted to the space-time description in
which the cohomology classes are required to be harmonic up the fibres
of the fibration $p:\PT\to\M$; this reduces $(a,g)$ to the forms given
in equations (\ref{harmgauge}).  In this latter gauge the quantum
field theory will be equivalent to the standard space-time formulation
because the Faddeev-Popov determinants will be independent of $A$ and
$G$.  The task then is to use BRST to see that the quantum theory is
the same in these two different gauges.  In particular, we would like
to see that all loops, and only Yang-Mills loops are obtained by some
suitable interpretation or generalisation of equation
(\ref{tsgenerate}) in which general algebraic curves of higher genus
contribute to $\CM^d_s$.
We will discuss this problem in a subsequent paper.

\section{Conformal gravity}\label{confgrav}

Berkovits and Witten (2004) have analyzed the twistor-string
formulation of conformal (super-)gravity.  In this section we give the
twistor construction, action and twistor-string reformulation for
conformal gravity.  This proceeds very much analogously to the
corresponding ideas for Yang-Mills and so we will sketch the ideas
relatively briefly in this section.

We take twistor space $\cPT$ now to be a manifold diffeomorphic to
$\R^4\times S^2$ endowed with an almost complex structure $\J$, i.e.,
$\J$ is an endomorphism of the real tangent bundle satisfying
$\J^2=-1$.  For the case of anti-self dual conformal gravity,
Berkovits \& Witten provide an analogue of the truncation of the
Chern-Simons action on super twistor space which is a functional of
the almost complex structure tensor $\J$, and a second tensor $k$.

As usual, one can use $\J$ to define subbundles $T^{(0,1)}$ and
$T^{(1,0)}$ of the complexified tangent bundle $T_\C$ as the $-i$ and
$+i$ eigenspaces of $\J:T_\C\to T_\C$ respectively and then define
correspondingly subbundles $\Omega^{(p,q)}$ of the bundles
$\Omega^{p+q}$ of complex differential forms.  Similarly, $\p$ and
$\dbar$-operators can be defined as the projection of the exterior
derivative $\d$ acting on sections of $\Omega^{p,q}$ onto
$\Omega^{(p+1,q)}$ and $\Omega^{p,q+1}$ respectively.  However, in
general, we will have that $N:=\dbar^2\in\Omega^{(0,2)}\otimes T^{(1,0)}$
does not vanish. 

In order for the ingredients to make contact with ordinary twistor
theory, we must require that $\J$ is chosen so that canonical bundle
$\Omega^{(3,0)}$ has Chern 
class $-4$ on the $S^2$ factors.

The tensor $k$ is a section of $\Omega^{(1,1)}\otimes \Omega^{(3,0)}$.  The
twistor 
space Lagrangian for the anti-self dual field is
$$
S[\J,k]=\int_{\cPT}(\dbar^2, k)
$$ where the pairing $(,)$ denotes both the contraction of the
holomorphic $T^{(1,0)}$ index of $N=\bar\p^2$ with the $\Omega^{(1,0)}$ index
of $k$ and the wedge product
of the antiholomorphic form indices with each-other and the
$\Omega^{(3,0)}$ indices.  It is easily seen that the
action is diffeomorphism invariant.  Furthermore we have
\begin{lemma}
The action $S[\J,k]$ is invariant under
$k\rightarrow k+ \dbar c$, where $c$ is a compactly supported section of
$\Omega^{(1,0)}\otimes\Omega^{(3,0)}$.
\end{lemma}
\Proof This follows from an identity obeyed by $N$ that arises as
follows.
 In general the exterior derivative maps
$$
\d:\Omega^{(p,q)}\rightarrow\Omega^{(p+2,q-1)}
\oplus\Omega^{(p+1,q)}\oplus\Omega^{(p,q+1)}\oplus\Omega^{(p-1,q+2)}
\, .
$$
The map $\d:\Omega^{(p,q)}\rightarrow\Omega^{(p+2,q-1)}$ is given
by contraction with the vector index of $-N$ and wedge product over the
form indices which can we write as $\alpha\rightarrow -N\hook\wedge
\alpha$.  The map $\d:\Omega^{(p,q)}\rightarrow\Omega^{(p-1,q+2)}$ is
similarly determined by $\bar N$.  It is a consequence of $\d^2=0$
that for $\alpha\in\Omega^{(1,0)}$,
$\dbar(N\hook\alpha)-N\hook\wedge\dbar\alpha=0$.  We can
therefore see that if $k=\dbar (c\otimes \nu)$, where $\nu$ is a
section of $\Omega^{(3,0)}$ and $c$ a section of $\Omega^{(1,0)}$,
then we have
$$
(N,\dbar (c\otimes \nu))=
(N\hook\wedge\dbar c)\wedge \nu + (N\hook c)\wedge\dbar\nu=\d(N\hook
c\wedge\nu) 
$$
and this implies the appropriate gauge invariance.$\Box$

\medskip

The field equations from this action are that $\dbar^2=0$, i.e., that
$\J$ should be integrable and $\dbar k=0$.  Given the gauge
invariance, $k$ defines an element of $H^1(\cPT,
\Omega^{(1,0)}\otimes\Omega^{(3,0)})$.  The standard
nonlinear-graviton construction, Penrose (1976), applied to $\cPT$
constructs a complex 4-manifold $\CM$ with holomorphic conformal
structure $[g]$ that has anti-self-dual Weyl curvature.  $\CM$ is the
space of rational curves in $\cPT$ with normal bundle
$\O(1)\oplus\O(1)$ (this requires either the existence of one rational
curve in $\cPT$ with normal bundel $\O(1)\oplus\O(1)$ or that $\J$ be
close to the standard complex structure on a neighbourhood of a line
in $\CP^3$).  The Penrose transform for $k$ leads to a self-dual
spinor $K_{A'B'C'D'}$ obeying the equation \be{linsdweyl}
(\nabla^{A'}_A\nabla^{B'}_B + \Phi^{A'B'}_{AB})K_{A'B'C'D'}=0\, , \ee
which are the linearised self-dual conformal gravity equations with
$K_{A'B'C'D'}$ playing the role of an infinitesimal self-dual Weyl
spinor on the anti-self-dual background.  We will see how this can be
done explicitly in a somewhat more general context later.  We note,
following Atiyah, Hitchin and Singer (1978), that $\CM$ admits a real
slice $M$ on which the conformal structure has Euclidean signature iff
$\cPT$ admits a conjugation $\hat{\;}:\cPT\to\cPT$ (i.e., it reverses the
sign of $\J$) with no fixed points.  The real space-time $M$ is then
the space of rational curves that are left invariant by the
conjugation.

Following Berkovits \& Witten, we note that the above action corresponds to the
space-time action 
$$
S_{\mbox{asd}}[g, K]=\int_M \psi^{A'B'C'D'}K_{A'B'C'D'}\d^4x\, .
$$
for anti-self-dual conformal gravity on space-time where $g$ is a
conformal structure and $\psi_{A'B'C'D'}$ is its self-dual Weyl
spinor.  The field equations implies the vanishing of
$\psi_{A'B'C'D'}$ and equation (\ref{linsdweyl}) for $K_{A'B'C'D'}$.

\subsection{The extension to non anti-self-dual fields}
This ASD space-time action can be extended to full
conformal gravity if we include the term
$$
I[g,K]=\frac\epsilon2\int_M K^{A'B'C'D'}K_{A'B'C'D'}\d^4x\, .
$$
We will reformulate this on twistor space in terms of $k$ to obtain
$I[k, \J]$.
We will take our integral to be that of a product of $k_1$ and $k_2$
over an 8-dimensional contour in $\cPT\times\cPT$.  We must first
develop the Penrose non-linear graviton construction in the case that the
complex structure $\J$ is not integrable in order to define the
ingredients that we will need.

We first introduce a conjugation
$\hat{\;}:\cPT\rightarrow \cPT$, $\hat{}\;^2=1$ that reverses $\J$,
i.e., $\hat{\;}^*\J=-\J$.  
There are two types of such conjugations normally employed in twistor
theory, 
depending on whether the conjugation has fixed points in twistor
space or not.  The latter case leads to Euclidean signature on
space-time and we will assume that to be the case hereon.
  
We now consider the moduli space $\CM$ of pseudo-holomorphic rational
curves in $\cPT$, i.e., the space of embedded $S^2$s in $\cPT$ in the
same topological class as the $S^2$ factors in $\cPT=\R^4\times S^2$,
such that $\J$ leaves the tangent space invariant inducing a complex
structure thereon.  Theorems in McDuff and Salamon (2004) imply that
$\CM$ exists and is 8-dimensional if $\J$ is close to the standard
complex structure on a neighbourhood of a line in $\CP^3$ (and we will
assume this to be the case hereon).  This
follows from the ellipticity of the equations defining such a
$\J$-holomorphic curve and the index theorem applied to its
linearization.  The conjugation $\hat{\;}$ induces a conjugation
$\hat{\;}:\CM\rightarrow \CM$, $\hat{}\;^2=1$ and we define $M$ to be
the (4-dimensional) fixed point set of $\hat{\;}$ on $\CM$.

We take $M$ to be our candidate space-time and we will have a
projection $p:\cPT\rightarrow M$ as a consequence of the fact that,
with our assumptions, there will be a unique rational curve in $\cPT$
through $Z$ and $\hat Z$.  The fibres of $p$ are, by construction, Riemann
spheres, $\CP^1$.

We define $\cT$ to be the total space of the line bundle
$\left(\Omega^{(3,0)}\right)^{1/4}$ (this 4th root exists as a
consequence of our assumptions on the topology of $\cPT$ and $\J$, in
particular that $\Omega^{(3,0)}$ has Chern class $-4$).  Since
$\Omega^{(3,0)}$ is an almost complex line bundle, its total space and
its powers are almost complex, so that $\cT$ has an almost complex
structure.  We denote the complex line bundles
$\left(\Omega^{(3,0)}\right)^{-n/4}$ by $\O(n)$.  On restriction to
each $\CP^1$ fibre, $\cT$ will be a line bundle of degree $-1$ and is
hence the tautological bundle on each $\CP^1$ fibre of $p$.  Let
$\tilde p:\cT\to M$ denote the projection induced by $p$.  The fibres
of $\tilde p$ minus the zero section are canonically identifiable with
the complement of the zero-section in a rank two vector bundle with
structure group $\SU(2)$ over $M$ and, with an abuse of notation, we
will think of $\cT$ as being this complex rank two vector bundle.
Introduce linear coordinates $\pi_{A'}$, $A'=0',1'$ on the fibres of
$\tilde p$.  Define the Euler homogeneity operator
$\Upsilon=\pi_{A'}\p/\p\pi_{A'}$.

We choose a frame for $\Omega^{(1,0)}(\cT)$ as follows.  Choose
$1$-forms $D\pi_{A'}$ in $\Omega^{(1,0)}(\cT)$ of homogeneity degree 1
in $\pi_{A'}$, $\Lie_\Upsilon D\pi_{A'}=D\pi_{A'}$, and so that on
restriction to the fibres of $\tilde p$, $D\pi_{A'}=\d\pi_{A'}$.  In
order to achieve this, in general $D\pi_{A'}$ will have to have
non-holomorphic dependence on $\pi_{A'}$.  The 1-form
$\D\pi:=\pi^{A'}\D\pi_{A'}$ descends to $\cPT$ to give a 1-form with
values in $\O(2)$.  At each point we can find a pair of complex
1-forms, $\theta^A$, $A=0,1$ homogeneous degree 1 in $\pi_{A'}$, such
that $\theta^A$ are orthogonal to the fibres of $p$ and to
$T^{(0,1)}$.  The condition that $\Omega^{(3,0)}=\O(-4)$ by definition
means that we have a canonical section $\Omega$ of
$\Omega^{(3,0)}\otimes\O(4)$.  Thus, since $\theta^A$ are sections of
$\O(1)\otimes\Omega^{(1,0)}$ we can also require
$\Omega=\theta^0\wedge\theta^1\wedge\pi_{A'}\d\pi^{A'}$.  Such
$\theta^A$ can be chosen to be global and non-vanishing on $\cPT$.
This gives our basis $\theta^\alpha=(\theta^A,D\pi_{A'})$ of
$\Omega^{(1,0)}$.

We can now study the Penrose transform of $k$ by setting
$$
k=(k^{A'}\wedge\D\pi_{A'}+k_A\wedge\theta^A)\otimes\Omega
$$
where $k^{A'}$ and $k_A$ are $(0,1)$-forms of homogeneity degree
$-5$.  Note that $\Upsilon\hook k=0$ so $k^{A'}\pi_{A'}=0$ so that
$k^{A'}=\pi^{A'}\varkappa$ for some $(0,1)$-form $\varkappa$ with
values in $\O(-6)$.  We can now finally define the indexed 2-form
$K^{A'}_{B'}$ on $M$ by
$$
K^{A'}_{B'}(x)
=\int_{L_x} \pi_{B'} k^{A'}\wedge\Omega\, =\int_{L_x} \pi_{B'}
\pi^{A'}\varkappa \wedge\Omega\, .
$$
Clearly $K^{A'}_{A'}=0$.  We then define
$$
I[k,\J]=\int_M K^{A'}_{B'}\wedge K^{B'}_{A'}\, .
$$
This can be expressed directly in terms of $k$ and 
$\varkappa$ as follows.  Let $\cPT\times_M\cPT$ be the 8-dimensional
space which fibres over $M$ with fibre $\CP^1\times\CP^1$, the
cartesian product of two copies of the fibre of $p:\cPT\to M$. This
has two projections $p_1$ and  $p_2$ onto $\cPT$, one on each factor. Let
$k_1=p_1^*k$ and $k_2=p_2^*k$ and similarly
$\varkappa_1=p_1^*\varkappa$ etc..  Our integral, then, is
$$
I[k,\J]=\int_{\cPT\times_M\cPT}
(\pi_1\cdot\pi_2)^2\varkappa_1\wedge\varkappa_2
\wedge\Omega_1\wedge\Omega_2\, .
$$
where $(\pi_{1A'},\pi_{2B'})$ are homogeneous coordinates on the
$\CP^1\times \CP^1$ fibres of $\cPT\times_M\cPT\to M$.  This integral
is invariant under $k\rightarrow k+\dbar l$ since this induces a
variation of the integrand in $I[k,\J]$ that is exact on the fibres of
$p_1$ and $p_2$.

\begin{propn}
  Solutions to the classical field equations up to diffeomorphism 
  arising from the action
  $S_T[\J,k]=S_{\mbox{asd}}[\J,k]-\frac\epsilon 2 I[k,\J]$ on twistor
  space are in one to one correspondence with solutions to the
  conformal gravity equations up to diffeomorphism.
\end{propn}

\Proof
We proceed as
before for Yang-Mills and focus on the field equation that arises from
varying $k$.  
Varying first with respect to $k_A$ we find
$$\dbar^2\hook\theta^A=0.$$ 
Varying  $k^{A'}$ (or equivalently $\varkappa$) we obtain:  
\be{confgrav1}
(\dbar^2\hook \D\pi
)|_Z=\int_{p^{-1}(p(Z))}
\pi^{B'}\pi_{A'}\pi^1_{B'}k_1^{A'}\wedge\Omega_1=
\pi_{A'}\pi^{B'}K_{B'}^{A'(0,2)}
\ee
where the subscript $(0,2)$ on a 2-form denotes projection onto the
$(0,2)$ part.  

In particular, the (0,2)-form part of $\dbar^2$ annihilates vertical
vectors.  The normal bundle to the fibre $L_x$ of $p$ over $x$ is therefore a
holomorphic vector bundle on $L_x$.  On each $L_x$,  $\theta^A$
can therefore be chosen uniquely up to a fibrewise global $\GL(2,\C)$
action on the $A$ index to be
holomorphic.  The topological assumption that the canonical bundle of
twistor space is $\O(-4)$ means that the normal bundle should have
first Chern class 2, so that it has generic splitting type
$\O(1)\oplus\O(1)$, and must be isomorphic to this with our assumption that
$\J$ is close to the standard one; $\theta^A$ can be defined to be
this isomorphism.  Since $\theta^A$ is global and holomorphic in
$\pi_{A'}$ with homogeneity degree 1, there exists 1-forms
$\theta^{AA'}$ on $M$ such that $\theta^A=\theta^{AA'}\pi_{A'}$.  This
yields a conformal structure
$$
\d s^2=\varepsilon_{AB}\varepsilon_{A'B'}\theta^{AA'}\theta^{BB'}\, .
$$
Similarly, $\D\pi_{A'}$ can be chosen to be holomorphic up the
fibres and chosen globally up to a freedom $\D\pi_{A'}\rightarrow
\D\pi_{A'}+\gamma_{AA'}\theta^A$ for some $\gamma_{AA'}$ that depends
only on $x$.  Equation (\ref{confgrav1}) implies the vanishing of 
the conformally invariant part of the torsion of the connection
determined by the horizontal subspaces defined by $\D\pi_{A'}$. 

We can now see that $M$ is a manifold with Riemmannian conformal
structure and that $\cPT\to M$ is its projective spin bundle with the
standard twistorial almost complex structure as in Atiyah Hitchin and
Singer (1978).  For this almost complex structure, it is standard that
$$ \dbar^2=
\pi_{A'}\psi^{A'}_{B'C'D'}\varepsilon_{CD}\theta^{CC'}\wedge\theta^{DD'}
\frac{\p}{\p\pi_{B'}}
$$
where $\psi_{A'B'C'D'}$ is the self-dual Weyl-spinor.  Thus the field
equation obtained by varying $k$ implies that
$$
\psi_{A'B'C'D'}=\epsilon K_{A'B'C'D'}
$$

We can now see that the action reduces to the space-time
action $S_{\mathrm{asd}}[g,K] + I[g,K]$ and so this action is in fact
equivalent to that for conformal gravity.$ \Box$

We note that in the above, the diffeomorphism freedom appropriate to
full twistor space is broken to the diffeomeorphism freedom on
space-time, together with the automorphisms of the bundle of self-dual
spinors.  In order to similarly reduce the gauge freedom for $k$, we
can require that, on each fibre of $p$, it is a harmonic representative of
the restriction of the cohomology class.  As before, an explicit
expression for $k$ can be given in this gauge that satisfies the field
equations arising from the twistor action when the conformal
structure has vanishing Bach tensor.


\subsection{Reformulation on supertwistor space}\label{susyconfgrav}
As in the Yang-Mills case, we can rewrite the $I[g,K]$ term in terms
of an integral over $M$ and the two copies of the fibre $L_x$ of
$\cPT\to M$.
\be{confgravIasD}
I[k,\J]=
\int_{\cPT\times_M\cPT} 
(\pi_1\cdot\pi_2)^4
\varkappa_1\wedge\varkappa_2
\wedge\D\pi_1\wedge\D\pi_2\wedge\d^4x
\ee

As before, we can introduce $N=4$ super-twistor space $\cPT_s$ and its
correspondence with super space-time, $M_s$ as follows.  Let $\psi_i$,
$i=1,\ldots, 4$ be anticommuting variables on the supersymmetric
twistor space $\cPT_s$ with values in $\O(1)$, and let $\theta_i^{A'}$
be the corresponding anti-commuting coordinates on the super
space-time $M_s$ with incidence relation
$\psi_i=\pi_{A'}\theta^{A'}_i$.  As before set
$\Psi=\psi_1\psi_2\psi_3\psi_4$ and
$\Omega^s=\Omega\wedge \d\psi_1\wedge\ldots\wedge\d\psi_4$.  Using
again 
the relation $\int \d^8\theta \;\Psi_1\Psi_2=(\pi_1\cdot\pi_2)^4$ we can
write
\begin{eqnarray}
I[k,\J]&=&\int_{\cPT_s\times_M\cPT_s} (\pi_1\cdot\pi_2)^{-2}(\Psi_1
\varkappa_1 )\wedge(\Psi_2\varkappa_2
)\wedge\Omega^s_1\wedge\Omega^s_2 \nonumber\\ 
&=&
\int_{M_s\times L(x,\theta)\times L(x,\theta)}
(\Psi_1\varkappa_1\wedge\D\pi_1) \wedge (\Psi_2\varkappa_2 \wedge
\D\pi_2)\wedge  \d^4x\wedge\d^8\theta\nonumber \\
&=&
\int_{M_s\times L(x,\theta)\times L(x,\theta)}
(\Psi_1 k)\wedge(\Psi_2 k)\wedge
\d^4x\wedge\d^8\theta\nonumber\\
&=&\int_{M_s} \d^4x\d^8\theta \left(\int_{L(x,\theta)}\Psi k\right)^2
\end{eqnarray}
where the second last identity follows simply from the fact that
$k|_{L(x,\theta)}=\varkappa\wedge\D\pi$ and we now think of
$k\in\Omega^{(1,0)}(-4)$ rather than as a 1-form with values in
$\Omega^{(3,0)}$.

\subsection{Twistor-string theory for conformal gravity}\label{tsconfgrav}
For simplicity we work with formal path integral formulae.
Following the logic of \S\ref{twistorstring} backwards now, we start
with the twistor version of the path integral for conformal gravity
and work towards a formulation along the lines of equation
(\ref{tsgenerate}).  

We have that the generating functional for conformal gravity
scattering in terms of the twistor Lagrangians is
$$
\cA[\J,k]=\int \D\tilde\J\, \D\tilde k\;
e^{S_{\mathrm{asd}}[\tilde\J,\tilde k]-\frac{\epsilon^2}{2} I[\tilde
    k,\tilde \J]}
$$
where again the path integral is over fields $\tilde \J$
and $\tilde k$ that are suitably aymptotic to $\J$ and $k$.

We can manipulate this as before to obtain:
\begin{eqnarray}
\cA[\J,k]&=&\int \D\tilde\J\, \D\tilde k\;
e^{S_{\mathrm{asd}}[\tilde \J,\tilde
    k]}
\sum_{d=0}^\infty\frac{\epsilon^{2d}I[\tilde k,\tilde
    \J]^d}{2^dd!} \nonumber \\
&=&
\int \D\tilde\J\, \D\tilde k\;
e^{S_{\mathrm{asd}}[\tilde \J,\tilde
    k]} 
\sum_{d=0}^\infty \frac{\epsilon^{2d}}{2^d d!}
\int_{M_s^d}\Pi_{r=1}^d \d^4x_r\d^8\theta_r
\left(\int_{L(x_r,\theta_r)}\Psi_r\tilde k_r\right)^2\nonumber \\ 
&=& 
\sum_{d=0}^\infty 
\int_{M_s^d}\Pi_{r=1}^d \d^4x_r\d^8\theta_r
\int \D\tilde\J\, \D\tilde k\;
e^{S_{\mathrm{asd}}[\tilde \J,\tilde
    k] + \epsilon\sum_{r=1}^d \int_{L(x_r,\theta_r)} \Psi \tilde k}
\end{eqnarray}
This yields the coupling of the D1 instantons, $L(x_r,\theta_r)$, to
the 1-form $\Psi k$ precisely as proposed in Berkovits and Witten
(2004).

\section*{References}
Atiyah, M.F., Hitchin, N.J.\ and Singer, I.M.\ (1978) Self-duality in
four dimensional Riemannian geometry, {\it  Proc. Roy Soc. Lond.},
{\bf A 362}, 425-61.

\smallskip

\noindent
Berkovits, N., and Witten, E. (2004) Conformal supergravity in
Twistor-String theory, arXiv: hep-th/0406051.

\smallskip

\noindent
Cachazo, F., and Svrcek, P. (2005) Lectures on twistor strings and
perturbative Yang-Mills theory, arXiv:hep-th/0504194.

\smallskip

\noindent
Cachazo, F., Svrcek, P. and Witten, E. (2004) MHV vertices and tree
amplitudes in gauge theory, JHEP, 0409:006, arXiv:hep-th/0403047.

\smallskip

\noindent
Chalmers, G., and Siegel, W. (1996) The self-dual sector of QCD amplitudes,
Phys. Rev. D54, 7628-33.
arXiv:hep-th/9606061.

\smallskip

\noindent
Faddeev, L., and Slavnov, A. (1991) Gauge fields: an introduction to
quantum theory, Frontiers in Physics, Perseus.

\smallskip

\noindent
Gukov, S., Motl, L., and  Nietzke, A., (2004) Equivalence of twistor
prescriptions for super Yang-Mills, arXiv:hep-th/0404085.

\smallskip

\noindent
Mason, L.J., Singer, M.A., and Woodhouse, N.M.J.\ (2002)
Tau functions, twistor theory and quantum field theory, Comm.\ Math.\
Phys.\, {\bf 230}, no.\ 3, 389-420, arXiv: math-ph/0105038.

\smallskip

\noindent
McDuff, D., and Salamon, D. (2004) J-homolmorphic curves and
symplectic topology, Colloquium publications {\bf 52}, AMS.

\smallskip

\noindent
Penrose, R. (1976) Nonlinear gravitons and curved twistor theory,
Gen. Rel. Grav., {\bf 7}, 31-52.

\smallskip

\noindent
Roiban, R., Spradlin, M., and Volovich, A. (2004) On the tree level
S-matrix for Yang-Mills theory, Phys. Rev. D70, 026009,
arXiv:hep-th/0403190.  See also: Roiban, R., Spradlin, M., and
Volovich, A. (2004) A googly amplitude from the B model on Twistor
space, JHEP 0404, 012, arXiv:hep-th/0402016 and Roiban, R., and
Volovich, A. (2004) All googly amplitudes from the B model in Twistor
space, Phys.\ Rev.\ Lett.\ {\bf 93}, 131602, arXiv:hep-th/0402121.

\smallskip

\noindent
Witten, E. (2004)  Perturbative gauge theory as a string theory in
twistor space, {\it Comm. Math. Phys.}, {\bf 252}, p189,
arXiv:hep-th/0312171. 

\smallskip

\noindent
Woodhouse, N.M.J (1985) Real methods in twistor theory, Class.\
Quant.\ Grav.\, {\bf 2}, 257-91.

\end{document}